\RequirePackage{amsmath}
\RequirePackage{flexisym}
\RequirePackage[superscript,biblabel,nomove]{cite}
\RequirePackage{graphicx}
\documentclass[12pt,aasms4]{article}
\usepackage{amsmath}
\usepackage{amssymb}
\usepackage{pbox}
\usepackage{indentfirst}
\usepackage[hyphens]{url}
\usepackage[bookmarks=false,hidelinks]{hyperref}
\usepackage[hyphenbreaks]{breakurl}
\usepackage{caption}
\usepackage{float}
\usepackage{threeparttable}
\usepackage{setspace}
\def\gsim{\;\rlap{\lower 2.5pt \hbox{$\sim$}}\raise 1.5pt\hbox{$>$}\;}
\def\lsim{\;\rlap{\lower 2.5pt  \hbox{$\sim$}}\raise 1.5pt\hbox{$<$}\;}
\newcommand\beq{\begin{equation}}
\newcommand\eeq{\end{equation}}

\begin{document}

\Large
\centerline{\bf A Lunar Backup Record of Humanity}

\medskip
\medskip
\normalsize
\centerline{ Carson Ezell$^{1\star}$, Alexandre Lazarian$^{2,3}$, \& Abraham Loeb$^1$}
\medskip
\medskip
\medskip
\centerline{\it $^1$Department of Astronomy, Harvard University}
\centerline{\it 60 Garden Street, Cambridge, MA 02138, USA}
\centerline{\it $^2$Department of Astronomy, University of Wisconsin-Madison}
\centerline{\it Madison, WI 53706, USA}
\centerline{\it $^3$Centro de Investigación en Astronomía, Universidad Bernardo O’Higgins}
\centerline{\it Santiago, General Gana 1760, 8370993, Chile}
\medskip
\medskip
\small
\centerline{$^{\star}$Correspondence to carson.ezell@cfa.harvard.edu}
\normalsize
\vskip 0.2in
\hrule
\vskip 0.2in

\begin{doublespace}
\abstract{The risk of a catastrophic or existential disaster for our civilization is increasing this century. A significant motivation for a  near-term space settlement is the opportunity to safeguard civilization in the event of a planetary-scale disaster. A catastrophic event could destroy the significant cultural, scientific, and technological progress on Earth. However, early space settlements can preserve records of human activity by maintaining a backup data storage system. The backup can also store information about the events leading up to the disaster. The system would improve the ability of early space settlers to recover our civilization after collapse. We show that advances in laser communications and data storage enable the development of a data storage system on the lunar surface with a sufficient uplink data rate and storage capacity to preserve valuable information about the achievements of our civilization and the chronology of the disaster.}


\section{Concept}

A global catastrophic risk (GCR) is a ``risk that might have the potential to inflict serious damage to human well-being on a global scale''. \cite{bostrom_global_2011} The most severe GCRs include weapons of mass destruction, artificial pandemics, supervolcanic eruptions, \cite{mani_global_2021} or asteroid impacts. \cite{westin_global_2020} Some of these risks also pose existential threats to humanity. \cite{bostrom_analyzing_2002} Furthermore, recent exoplanet data have suggested an abundance of potentially habitable planets, ranging from 0.03 to over 1 per Sun-like star. \cite{bryson_occurrence_2021, hsu_occurrence_2019} The absence of signatures of living civilizations on exoplanets \cite{seager_exoplanet_2010} might suggest that civilizations are fragile. 

The National Aeronautics and Space Administration (NASA) plans to build its Artemis Base Camp on the lunar surface by the end of the current decade, \cite{noauthor_nasas_2020} and China is similarly planning a research station on the lunar South Pole. \cite{zou_tentative_2018} SpaceX has designed a technical roadmap for the  settlement of Mars. \cite{musk_making_2017} We can equip early space settlers with a record of human activity to improve the prospects of civilization recovery if a terrestrial catastrophe occurs. We propose a resilient, large-scale, lunar data storage system that utilizes laser communications for receiving updated data about ongoing human activity. 

The system would store the detailed records of our cultural, scientific, technological, and ecological information. It would also contain documents and imagery of mistakes and events leading up to the catastrophic disaster. Much like the black box of an airplane, the backup system would preserve the history of humanity after a catastrophe. 

Before being transported to the lunar surface, the system could be loaded with existing information about human civilization. Once on the moon, laser communications could be used to transfer data to the system at high rates. The system would store enough information to allow early space settlers to recover civilization, both socially and biologically. Furthermore, stored genetic information about other known species would allow our ecosystem to be reconstructed. 

The stored data should minimize biases and accurately portray human activities and their consequences. The Voyager 1 and Voyager 2 probes, launched by NASA in 1977, each carry a Golden Record that shares information about human civilization with any extraterrestrial civilization that may be encountered by them in the far future ({Source: \href{https://voyager.jpl.nasa.gov/golden-record/}{https://voyager.jpl.nasa.gov/golden-record/}} (accessed on 29 September 2022)).  While the Golden Record was designed to represent human civilization, the content curation process required making decisions about how we represent ourselves and who chooses, leading to selection biases.\cite{schmitt_archiving_2017} A lunar data backup would be more comprehensive than the Golden Record, but the sources from where data are drawn can also have selection biases. Actively sourcing records that are not already compiled in accessible databases, including previously unwritten descriptions of technical processes or cultural knowledge from underrepresented groups, would be a necessary component of the data curation process. Social scientists and other experts in inter-cultural communication are more well equipped than technical experts to ensure a selection of content that represents all of humanity and adequately portrays the information we wish to share, suggesting the need for their involvement in the content curation process. \cite{traphagan_should_2021} In some cases, the preservation of physical artifacts---such as crop samples \cite{westengen_global_2013}---may also be necessary to best represent our civilization and enable recovery.

Preserving records of the negative aspects of our civilization, including conflict and the negligence of catastrophic risks, could guide a future civilization to avoid making similar mistakes. If adequate knowledge is preserved, especially details about the chronology of the disaster itself, the survivors of the catastrophic event and future generations would have a deeper concern for existential and catastrophic risks than our current civilization. Future generations can then recover positive aspects of civilization while exercising caution with respect to catastrophic or existential risks by engaging in differential technological development.\cite{bostrom_analyzing_2002} The civilization that recovers may have a higher probability of overcoming global challenges and existential risks than our current civilization.

\section{Challenges}

Projects for the long-term preservation of data on Earth have been created in many locations that may be safe in a disaster, such as in mines, on mountains, and in the Arctic.\cite{guzman_extremely_2016} However, most existing projects aim to preserve information for thousands to millions of years, and they are limited by the lack of long-duration storage capabilities. \cite{guzman_extremely_2016} For example, the Memory of Mankind project is building a large, long-term storage facility in a saltmine that uses ceramic tablets as data carriers because their durability significantly exceeds that of digital systems (Source: \href{https://www.memory-of-mankind.com/}{https://www.memory-of-mankind.com/} (accessed on 22 October 2022)).

Another limitation of terrestrial data backup systems is that even remote locations on Earth may not be safe in the event of a catastrophic disaster. In addition, if survivors of the disaster are not aware of the backup storage systems, the systems may not be discovered for a long time. A disaster that leaves the Earth largely uninhabitable may also mean survivors in a self-sufficient lunar settlement would be better equipped to recover civilization, possibly on another celestial body. 

Previous work has described possibilities for long-duration storage systems on the moon which can survive catastrophic disasters and last millions of years, but the challenges include developing long-term storage solutions and communicating the information to future generations on Earth. \cite{turchin_surviving_2018} Our proposal follows the trend toward the establishment of base camps on the lunar surface itself. By preserving data for humans already on the lunar surface, we eliminate the need for very long duration storage solutions or the communication of information back to survivors on Earth.

Our proposal involves the technical challenges related to data storage and data transmission capacity. The overall quantity of data in the world already exceeds the data storage capacity significantly, and transmitting and storing data on the lunar surface will be more constrained than terrestrial data storage. The total volume of data created or copied worldwide was estimated at $6.4 \times 10^{22}$ bytes as of 2020, \cite{idc_worldwide_2021-1} and it is expected to increase to $1.75 \times 10^{23}$ bytes by 2025. \cite{reinsel_digitization_2018} However, the global data storage capacity was only about $6.7 \times 10^{21}$ bytes in 2021. \cite{idc_worldwide_2021} Although this demonstrates significant growth from a total storage capacity of $2.9 \times 10^{20}$ bytes in 2007, \cite{hilbert_worlds_2011} it is still an order of magnitude less than the total data. 

The urgency of GCRs suggests that we should include the backup system in our earliest space settlement plans, limiting stored and transmitted data to the most important information. We suggest storing books, academic research, movies, earth imagery, and genetic information from known species. The existing databases may provide large quantities of accessible information that would be useful for civilizational recovery (see Table \ref{tab1}). Ongoing content curation processes can contribute further records to the system over time to promote the representation of all of humanity and ensure enough data are preserved to enable a recovery. As the system ages or our storage capabilities improve, the system can be upgraded by human or robotic operators on the lunar surface.

\begin{table}[H]
\caption{Initial stored data and data rate based on current information stored in existing databases.}
\label{tab1}%
\setlength{\tabcolsep}{2.58mm}\begin{tabular}{@{}| l | l | l | l |@{}}
\hline
\textbf{Category} & \textbf{Quantity}  & \textbf{Stored Data (Bytes) {$^{\text{a}}$}} & \textbf{Data Rate (Bytes/yr) {$^{\text{b}}$}}\\
\hline
Books    & $3 \times 10^7$   & $6 \times 10^{13}$  & $6.7 \times 10^{12}$  \\
Journal Articles    & $8.45 \times 10^7$   & $8.45 \times 10^{13}$  & $4 \times 10^{12}$  \\
Films    & $5.4 \times 10^5$   & $1.08 \times 10^{15}$  & $3 \times 10^{13}$   \\
Genetic Information {$^{\text{c}}$} & N/A & $8.7 \times 10^{15}$ & N/A \\
Earth Imagery {$^{\text{d}}$}    & N/A  & $7.5 \times 10^{14}$  & $9 \times 10^{15}$   \\
\hline
\end{tabular}

 \pbox{\columnwidth}{\footnotesize $^{\text{a}}$ {Sources: ISBNdb (\url{https://isbndb.com/news} (accessed on 5 September 2022)), IMDb (\url{https://www.imdb.com/pressroom/stats/} (accessed on 5 September 2022)),  International Association of Scientific, Technical and Medical Publishers \cite{johnson_stm_2018}, Planet Labs (\url{https://www.planet.com/faqs/} (accessed on 5 September 2022)). }
     $^{\text{b}}$ {We estimated mean file sizes of books (2 mb), journal articles (1 mb), and films (2 gb).}
     $^{\text{c}}$ {We consider the number of catalogued species, \cite{mora_how_2011} and we conservatively estimate the mean genome size as the human genome size, \cite{venter_sequence_2001} multiplying by a factor of 10 for genetic diversity.}
     $^{\text{d}}$ {We consider collecting earth imagery data from a single source and storing only the previous month, limiting the total stored data}.}

   
\end{table}

\subsection{Storage Capacity}

The system must have sufficient capacity to store the necessary existing and future data. Let $C_i$ be a category of information stored (e.g., books, movies, journal articles, etc.) where $1 <i<n$, let $q_i$ be the existing quantity of bytes of data associated with category $i$, and let $r_i$ be the growth rate of the total quantity of data (i.e., data rate/stored data) for category $i$. Some categories of data demonstrate exponential growth, which is enabled by the accelerating growth in data creation and collection in the era of big data. \cite{kaisler_big_2013} However, some categories may be better fit by a linear growth model because rates of knowledge creation are constrained. Hence, we let $a_i$ be an indicator variable where $a_i = 1$ if the data for category $i$ fit an exponential growth model and $a_i = 0$ if the data fit a linear growth model. We assume that the system must have sufficient capacity to store all initial and future data until its capacity is increased. Then, the necessary storage capacity of the system in bytes, $S$, is, \begin{equation} S = \sum\limits_{i=1}^{n}q_i(a_ie^{r_id} + (1-a_i)(r_id + 1)) \text{ ,}\end{equation} where $d$ is the duration of time, in years, before the system is replaced or repaired.

Historically, space-based data systems (e.g., satellites) have served as data relays rather than large data storage centers. \cite{huang_envisioned_2018} Because of the adverse effects of the harsh conditions in outer space on equipment (e.g., radiation and extreme temperatures), shielding systems and cooling equipment would need to be developed before large-scale data storage would be feasible. Such systems are already being considered by private organizations, including Cloud Constellation and its satellite-based cloud infrastructure design known as Space-Belt.\cite{huang_envisioned_2018} The establishment of permanent operations on celestial bodies, such as the Artemis Base Camp, will also increase the demand for space-based storage solutions and accelerate the  technological development.

\subsection{Data Transfer Rate}

New data about ongoing human activity could be continuously transferred to the system through laser communications from Earth. The data rate of sending new data to the system must exceed the rate of new data production (See Figure 1). Let $r_U$ be the data rate of uplink communications to the system. Then, the necessary data rate, in bytes per year, is \begin{equation} r_U = \sum\limits_{i=1}^{n}q_ir_i(a_ie^{r_id} -a_i + 1) \text{ .}\end{equation} 

Laser communications can be used to transmit data from ground stations to the system at significantly higher rates than radiofrequency waves. \cite{toyoshima_ground--satellite_2008} However, the bandwidth of ground-to-space laser communications is limited by the scintillation caused by optical turbulence, atmospheric absorption, beam scattering, and the need for a clear line-of-sight path. \cite{majumdar_free-space_2005, majumdar_introduction_2008} Maintaining multiple ground stations to transmit data can mitigate the negative effects of cloud cover on transmission. A constant data stream at $622$ Mbps from a ground station to the lunar surface—a rate which has been achieved by NASA's Lunar Laser Communications Demonstration (LLCD) \cite{boroson_overview_2014,devoe_optical_2017}—would be sufficient to transmit over $10^{15}$ 
bytes/yr. This is sufficient to transmit all of the aforementioned literature given the current growth rates. Near-term improvements in data rates would enable the transmission of enough data to include satellite imagery produced by the Dove and SkySat constellations operated by Planet Labs ({Source: \href{https://www.planet.com/faqs/}{https://www.planet.com/faqs/} (accessed on 31 August 2022)). We assume that redundant satellite imagery or higher-resolution imagery from future earth observation (EO) satellites is not necessary for transmission and storage.}).

Laser communication data rates have historically demonstrated exponential growth (See Figure 2), and a previous analysis suggests the U.S. Department of Defense requires an increase in bits per second by an order of magnitude every nine years. \cite{dmytryszyn_lasers_2021} NASA's Artemis Program is also utilizing laser communications, and its Orion EM-2 Optical Communications Terminal (O2O) will be the first usage of optical communications for human spaceflight.\cite{seas_optical_2018}  Future systems can maintain the growth by leveraging new modulation schemes. The existing techniques, such as pulse position modulation (PPM) and differential phase-shift keying (DPSK) modulation, \cite{xu_differential_2004} have enabled our current capabilities, and the usage of wavelength division multiplexing (WDM) may enable even greater data rates.\cite{toyoshima_recent_2021} Terrestrial fibers have already achieved exa-bps ($10^{18}$ bits per second) using WDM,\cite{toyoshima_current_2015} and uplink transmissions to satellites may experience similar increases in capabilities.

\begin{figure}[H]%
\centering
\includegraphics[width=0.65\textwidth]{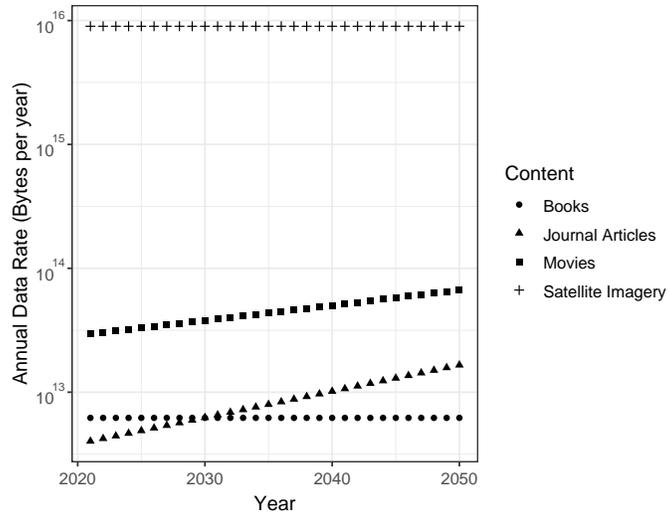}
\caption{Projected data rate for new content given historical growth rates.}\label{fig2}

\end{figure}

\begin{figure}[H]%
\centering
\includegraphics[width=0.65\textwidth]{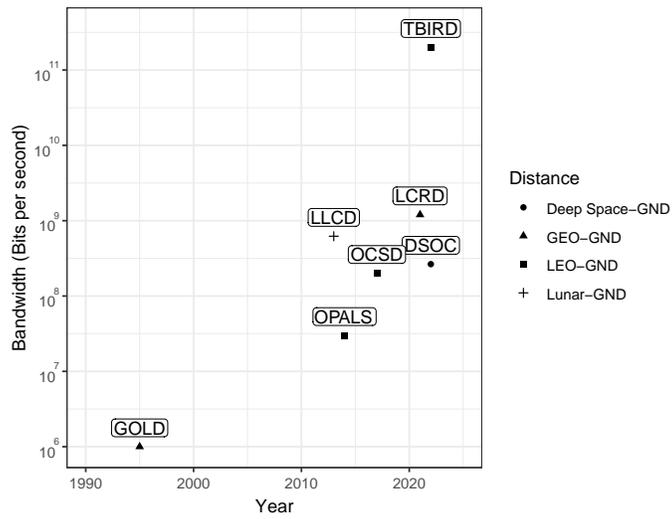}
\caption{Maximum bandwidth of previous and planned NASA laser communications demonstrations. \cite{toyoshima_recent_2021}}\label{fig1}
\end{figure}
\vspace{-6pt}

\section{Conclusion}

Lunar exploration in the near future suggests the present is an optimal time to consider the initial infrastructure for the first space bases. Early space settlement designs should strongly consider the importance of catastrophic risk mitigation, \cite{gottlieb_space_2019} including by developing the described data storage system. 

Whether the system enables civilization recovery still depends upon the self-sufficiency of early space settlements or how effective they are as refuges from catastrophic disasters.\cite{beckstead_how_2015} Self-sufficiency can be achieved sooner if existential risk mitigation is a motivating force for space development. Ensuring space settlers have the capacity to begin the civilizational recovery process—including access to spacecraft, energy, manufacturing, and critical resources—can further increase recovery prospects. The data system's chance of surviving a catastrophe may be increased if the project is a shared, international undertaking to reduce its vulnerability in a great power conflict. International, inter-generational coordination would also enable further initiatives to improve the prospects of the long-term future.

\end{doublespace}

\vskip 0.45in
\hrule
\vskip 0.15in

\small
\noindent



\bibliography{main}

\begin{thebibliography}{10}
\expandafter\ifx\csname url\endcsname\relax
  \def\url#1{\texttt{#1}}\fi
\expandafter\ifx\csname urlprefix\endcsname\relax\def\urlprefix{URL }\fi
\providecommand{\bibinfo}[2]{#2}
\providecommand{\eprint}[2][]{\url{#2}}

\bibitem{bostrom_global_2011}
\bibinfo{author}{Bostrom, N.} \& \bibinfo{author}{Cirkovic, M.~M.}
\newblock \emph{\bibinfo{title}{Global {Catastrophic} {Risks}}}
  (\bibinfo{publisher}{Oxford University Press}, \bibinfo{address}{Oxford, New
  York}, \bibinfo{year}{2011}).

\bibitem{mani_global_2021}
\bibinfo{author}{Mani, L.}, \bibinfo{author}{Tzachor, A.} \&
  \bibinfo{author}{Cole, P.}
\newblock \bibinfo{title}{Global catastrophic risk from lower magnitude
  volcanic eruptions}.
\newblock \emph{\bibinfo{journal}{Nature Communications}}
  \textbf{\bibinfo{volume}{12}}, \bibinfo{pages}{4756} (\bibinfo{year}{2021}).
\newblock \urlprefix\url{https://www.nature.com/articles/s41467-021-25021-8}.
\newblock \bibinfo{note}{Number: 1 Publisher: Nature Publishing Group}.

\bibitem{westin_global_2020}
\bibinfo{author}{Westin, U.}, \bibinfo{author}{Ingdahl, W.},
  \bibinfo{author}{Wariaro, V.} \& \bibinfo{author}{Shandwick, W.}
\newblock \bibinfo{title}{Global {Catastrophic} {Risks} 2020}
  (\bibinfo{year}{2020}).
\newblock
  \urlprefix\url{https://globalchallenges.org/wp-content/uploads/Global-Catastrophic-Risks-2020-Annual-Report.pdf}.

\bibitem{bostrom_analyzing_2002}
\bibinfo{author}{Bostrom, N.}
\newblock \bibinfo{title}{Analyzing {Human} {Extinction} {Scenarios} and
  {Related} {Hazards}}.
\newblock \emph{\bibinfo{journal}{Journal of Evolution and Technology}}
  \textbf{\bibinfo{volume}{9}} (\bibinfo{year}{2002}).

\bibitem{bryson_occurrence_2021}
\bibinfo{author}{Bryson, S.} \emph{et~al.}
\newblock \bibinfo{title}{The {Occurrence} of {Rocky} {Habitable}-zone
  {Planets} around {Solar}-like {Stars} from {Kepler} {Data}}.
\newblock \emph{\bibinfo{journal}{The Astronomical Journal}}
  \textbf{\bibinfo{volume}{161}}, \bibinfo{pages}{36} (\bibinfo{year}{2021}).
\newblock
  \urlprefix\url{https://ui.adsabs.harvard.edu/abs/2021AJ....161...36B}.
\newblock \bibinfo{note}{ADS Bibcode: 2021AJ....161...36B}.

\bibitem{hsu_occurrence_2019}
\bibinfo{author}{Hsu, D.~C.}, \bibinfo{author}{Ford, E.~B.},
  \bibinfo{author}{Ragozzine, D.} \& \bibinfo{author}{Ashby, K.}
\newblock \bibinfo{title}{Occurrence {Rates} of {Planets} {Orbiting} {FGK}
  {Stars}: {Combining} {Kepler} {DR25}, {Gaia} {DR2}, and {Bayesian}
  {Inference}}.
\newblock \emph{\bibinfo{journal}{The Astronomical Journal}}
  \textbf{\bibinfo{volume}{158}}, \bibinfo{pages}{109} (\bibinfo{year}{2019}).
\newblock
  \urlprefix\url{https://ui.adsabs.harvard.edu/abs/2019AJ....158..109H}.
\newblock \bibinfo{note}{ADS Bibcode: 2019AJ....158..109H}.

\bibitem{seager_exoplanet_2010}
\bibinfo{author}{Seager, S.} \& \bibinfo{author}{Deming, D.}
\newblock \bibinfo{title}{Exoplanet {Atmospheres}}.
\newblock \emph{\bibinfo{journal}{Annual Review of Astronomy and Astrophysics,
  vol. 48, p.631-672}} \textbf{\bibinfo{volume}{48}}, \bibinfo{pages}{631}
  (\bibinfo{year}{2010}).
\newblock
  \urlprefix\url{https://ui.adsabs.harvard.edu/abs/2010ARA%26A..48..631S/abstract}.

\bibitem{noauthor_nasas_2020}
\bibinfo{title}{{NASA}'s {Plan} for {Sustained} {Lunar} {Exploration} and
  {Development}}.
\newblock \bibinfo{type}{Tech. Rep.}, \bibinfo{institution}{National
  Aeronautics and Space Administration} (\bibinfo{year}{2020}).
\newblock
  \urlprefix\url{https://www.nasa.gov/sites/default/files/atoms/files/a_sustained_lunar_presence_nspc_report4220final.pdf}.

\bibitem{zou_tentative_2018}
\bibinfo{author}{Zou, Y.}, \bibinfo{author}{Xu, L.} \& \bibinfo{author}{Jia,
  Y.}
\newblock \bibinfo{title}{A {Tentative} {Plan} of {China} to {Establish} a
  {Lunar} {Research} {Station} in the {Next} {Ten} {Years}}
  \textbf{\bibinfo{volume}{42}}, \bibinfo{pages}{B3.1--34--18}
  (\bibinfo{year}{2018}).
\newblock
  \urlprefix\url{https://ui.adsabs.harvard.edu/abs/2018cosp...42E3886Z}.
\newblock \bibinfo{note}{Conference Name: 42nd COSPAR Scientific Assembly ADS
  Bibcode: 2018cosp...42E3886Z}.

\bibitem{musk_making_2017}
\bibinfo{author}{Musk, E.}
\newblock \bibinfo{title}{Making {Humans} a {Multi}-{Planetary} {Species}}.
\newblock \emph{\bibinfo{journal}{New Space}} \textbf{\bibinfo{volume}{5}},
  \bibinfo{pages}{46--61} (\bibinfo{year}{2017}).
\newblock
  \urlprefix\url{https://www-liebertpub-com.ezp-prod1.hul.harvard.edu/doi/10.1089/space.2017.29009.emu}.
\newblock \bibinfo{note}{Publisher: Mary Ann Liebert, Inc., publishers}.

\bibitem{schmitt_archiving_2017}
\bibinfo{author}{Schmitt, R.~M.}
\newblock \emph{\bibinfo{title}{Archiving "{The} {Best} of {Ourselves}" on the
  {Voyager} {Golden} {Record}: {Rhetorics} of the {Frontier}, {Memory}, and
  {Technology}}}.
\newblock Master's thesis, \bibinfo{school}{University of Colorado at Boulder},
  \bibinfo{address}{United States -- Colorado} (\bibinfo{year}{2017}).
\newblock
  \urlprefix\url{https://www.proquest.com/docview/1916508295/abstract/71D961D436FD4E56PQ/1}.
\newblock \bibinfo{note}{ISBN: 9781369865141}.

\bibitem{traphagan_should_2021}
\bibinfo{author}{Traphagan, J.~W.}
\newblock \bibinfo{title}{Should {We} {Lie} to {Extraterrestrials}? {A}
  {Critique} of the {Voyager} {Golden} {Records}}.
\newblock \emph{\bibinfo{journal}{Space Policy}} \textbf{\bibinfo{volume}{57}},
  \bibinfo{pages}{101440} (\bibinfo{year}{2021}).
\newblock
  \urlprefix\url{https://www.sciencedirect.com/science/article/pii/S0265964621000321}.

\bibitem{westengen_global_2013}
\bibinfo{author}{Westengen, O.~T.}, \bibinfo{author}{Jeppson, S.} \&
  \bibinfo{author}{Guarino, L.}
\newblock \bibinfo{title}{Global {Ex}-{Situ} {Crop} {Diversity} {Conservation}
  and the {Svalbard} {Global} {Seed} {Vault}: {Assessing} the {Current}
  {Status}}.
\newblock \emph{\bibinfo{journal}{PLOS ONE}} \textbf{\bibinfo{volume}{8}},
  \bibinfo{pages}{e64146} (\bibinfo{year}{2013}).
\newblock
  \urlprefix\url{https://journals.plos.org/plosone/article?id=10.1371/journal.pone.0064146}.
\newblock \bibinfo{note}{Publisher: Public Library of Science}.

\bibitem{guzman_extremely_2016}
\bibinfo{author}{Guzman, M.}, \bibinfo{author}{Hein, A.~M.} \&
  \bibinfo{author}{Welch, C.}
\newblock \bibinfo{title}{Extremely {Long}-{Duration} {Storage} {Concepts} for
  {Space}}.
\newblock \emph{\bibinfo{journal}{Acta Astronautica}}
  \textbf{\bibinfo{volume}{130}} (\bibinfo{year}{2016}).

\bibitem{turchin_surviving_2018}
\bibinfo{author}{Turchin, A.} \& \bibinfo{author}{Denkenberger, D.}
\newblock \bibinfo{title}{Surviving global risks through the preservation of
  humanity's data on the {Moon}}.
\newblock \emph{\bibinfo{journal}{Acta Astronautica}}
  \textbf{\bibinfo{volume}{146}} (\bibinfo{year}{2018}).

\bibitem{idc_worldwide_2021-1}
\bibinfo{author}{IDC}.
\newblock \bibinfo{title}{Worldwide {Global} {DataSphere} {Forecast},
  2021–2025: {The} {World} {Keeps} {Creating} {More} {Data} — {Now}, {What}
  {Do} {We} {Do} with {It} {All}?}
\newblock \bibinfo{type}{Tech. Rep.} \bibinfo{number}{\#US46410421},
  \bibinfo{institution}{IDC} (\bibinfo{year}{2021}).

\bibitem{reinsel_digitization_2018}
\bibinfo{author}{Reinsel, D.}, \bibinfo{author}{Gantz, J.} \&
  \bibinfo{author}{Rydning, J.}
\newblock \bibinfo{title}{The {Digitization} of the {World} {From} {Edge} to
  {Core}}.
\newblock \bibinfo{type}{{IDC} {White} {Paper}} \bibinfo{number}{US44413318},
  \bibinfo{institution}{Seagate} (\bibinfo{year}{2018}).
\newblock
  \urlprefix\url{https://www.seagate.com/files/www-content/our-story/trends/files/idc-seagate-dataage-whitepaper.pdf}.

\bibitem{idc_worldwide_2021}
\bibinfo{author}{IDC}.
\newblock \bibinfo{title}{Worldwide {Global} {StorageSphere} {Forecast},
  2021–2025: {To} {Save} or {Not} to {Save} {Data}, {That} {Is} the
  {Question}}.
\newblock \bibinfo{type}{Tech. Rep.} \bibinfo{number}{US47509621},
  \bibinfo{institution}{IDC} (\bibinfo{year}{2021}).

\bibitem{hilbert_worlds_2011}
\bibinfo{author}{Hilbert, M.} \& \bibinfo{author}{López, P.}
\newblock \bibinfo{title}{The {World}’s {Technological} {Capacity} to
  {Store}, {Communicate}, and {Compute} {Information}}.
\newblock \emph{\bibinfo{journal}{Science}} \textbf{\bibinfo{volume}{332}},
  \bibinfo{pages}{60--65} (\bibinfo{year}{2011}).
\newblock
  \urlprefix\url{https://www.science.org/doi/abs/10.1126/science.1200970}.
\newblock \bibinfo{note}{Publisher: American Association for the Advancement of
  Science}.

\bibitem{johnson_stm_2018}
\bibinfo{author}{Johnson, R.}, \bibinfo{author}{Watkinson, A.} \&
  \bibinfo{author}{Mabe, M.}
\newblock \bibinfo{title}{The {STM} {Report}: {An} overview of scientific and
  scholarly publishing} (\bibinfo{year}{2018}).
\newblock
  \urlprefix\url{https://www.stm-assoc.org/2018_10_04_STM_Report_2018.pdf}.

\bibitem{mora_how_2011}
\bibinfo{author}{Mora, C.}, \bibinfo{author}{Tittensor, D.~P.},
  \bibinfo{author}{Adl, S.}, \bibinfo{author}{Simpson, A. G.~B.} \&
  \bibinfo{author}{Worm, B.}
\newblock \bibinfo{title}{How {Many} {Species} {Are} {There} on {Earth} and in
  the {Ocean}?}
\newblock \emph{\bibinfo{journal}{PLOS Biology}} \textbf{\bibinfo{volume}{9}},
  \bibinfo{pages}{e1001127} (\bibinfo{year}{2011}).
\newblock
  \urlprefix\url{https://journals.plos.org/plosbiology/article?id=10.1371/journal.pbio.1001127}.
\newblock \bibinfo{note}{Publisher: Public Library of Science}.

\bibitem{venter_sequence_2001}
\bibinfo{author}{Venter, J.~C.} \emph{et~al.}
\newblock \bibinfo{title}{The {Sequence} of the {Human} {Genome}}.
\newblock \emph{\bibinfo{journal}{Science}} \textbf{\bibinfo{volume}{291}},
  \bibinfo{pages}{1304--1351} (\bibinfo{year}{2001}).
\newblock
  \urlprefix\url{https://www.science.org/doi/full/10.1126/science.1058040}.
\newblock \bibinfo{note}{Publisher: American Association for the Advancement of
  Science}.

\bibitem{kaisler_big_2013}
\bibinfo{author}{Kaisler, S.}, \bibinfo{author}{Armour, F.},
  \bibinfo{author}{Espinosa, J.~A.} \& \bibinfo{author}{Money, W.}
\newblock \bibinfo{title}{Big {Data}: {Issues} and {Challenges} {Moving}
  {Forward}}.
\newblock In \emph{\bibinfo{booktitle}{2013 46th {Hawaii} {International}
  {Conference} on {System} {Sciences}}}, \bibinfo{pages}{995--1004}
  (\bibinfo{year}{2013}).
\newblock \bibinfo{note}{ISSN: 1530-1605}.

\bibitem{huang_envisioned_2018}
\bibinfo{author}{Huang, H.}, \bibinfo{author}{Guo, S.} \&
  \bibinfo{author}{Wang, K.}
\newblock \bibinfo{title}{Envisioned {Wireless} {Big} {Data} {Storage} for
  {Low}-{Earth}-{Orbit} {Satellite}-{Based} {Cloud}}.
\newblock \emph{\bibinfo{journal}{IEEE Wireless Communications}}
  \textbf{\bibinfo{volume}{25}}, \bibinfo{pages}{26--31}
  (\bibinfo{year}{2018}).
\newblock \bibinfo{note}{Conference Name: IEEE Wireless Communications}.

\bibitem{toyoshima_ground--satellite_2008}
\bibinfo{author}{Toyoshima, M.} \emph{et~al.}
\newblock \bibinfo{title}{Ground-to-satellite laser communication experiments}.
\newblock \emph{\bibinfo{journal}{IEEE Aerospace and Electronic Systems
  Magazine}} \textbf{\bibinfo{volume}{23}}, \bibinfo{pages}{10--18}
  (\bibinfo{year}{2008}).
\newblock \bibinfo{note}{Conference Name: IEEE Aerospace and Electronic Systems
  Magazine}.

\bibitem{majumdar_free-space_2005}
\bibinfo{author}{Majumdar, A.~K.}
\newblock \bibinfo{title}{Free-space laser communication performance in the
  atmospheric channel}.
\newblock \emph{\bibinfo{journal}{Journal of Optical and Fiber Communications
  Reports}} \textbf{\bibinfo{volume}{2}}, \bibinfo{pages}{345--396}
  (\bibinfo{year}{2005}).
\newblock \urlprefix\url{https://doi.org/10.1007/s10297-005-0054-0}.

\bibitem{majumdar_introduction_2008}
\bibinfo{author}{Majumdar, A.~K.}
\newblock \bibinfo{title}{Introduction}.
\newblock In \bibinfo{editor}{Majumdar, A.~K.} \& \bibinfo{editor}{Ricklin,
  J.~C.} (eds.) \emph{\bibinfo{booktitle}{Free-{Space} {Laser}
  {Communications}: {Principles} and {Advances}}}, Optical and {Fiber}
  {Communications} {Reports}, \bibinfo{pages}{1--8}
  (\bibinfo{publisher}{Springer}, \bibinfo{address}{New York, NY},
  \bibinfo{year}{2008}).
\newblock \urlprefix\url{https://doi.org/10.1007/978-0-387-28677-8_1}.

\bibitem{boroson_overview_2014}
\bibinfo{author}{Boroson, D.~M.} \emph{et~al.}
\newblock \bibinfo{title}{Overview and results of the {Lunar} {Laser}
  {Communication} {Demonstration}}.
\newblock In \emph{\bibinfo{booktitle}{Free-{Space} {Laser} {Communication} and
  {Atmospheric} {Propagation} {XXVI}}}, vol. \bibinfo{volume}{8971},
  \bibinfo{pages}{213--223} (\bibinfo{publisher}{SPIE}, \bibinfo{year}{2014}).
\newblock
  \urlprefix\url{https://www.spiedigitallibrary.org/conference-proceedings-of-spie/8971/89710S/Overview-and-results-of-the-Lunar-Laser-Communication-Demonstration/10.1117/12.2045508.full}.

\bibitem{devoe_optical_2017}
\bibinfo{author}{DeVoe, C.~E.} \emph{et~al.}
\newblock \bibinfo{title}{Optical overview and qualification of the {LLCD}
  space terminal}.
\newblock In \emph{\bibinfo{booktitle}{International {Conference} on {Space}
  {Optics} — {ICSO} 2014}}, vol. \bibinfo{volume}{10563},
  \bibinfo{pages}{115--123} (\bibinfo{publisher}{SPIE}, \bibinfo{year}{2017}).
\newblock
  \urlprefix\url{https://www.spiedigitallibrary.org/conference-proceedings-of-spie/10563/105630F/Optical-overview-and-qualification-of-the-LLCD-space-terminal/10.1117/12.2304194.full}.

\bibitem{dmytryszyn_lasers_2021}
\bibinfo{author}{Dmytryszyn, M.}, \bibinfo{author}{Crook, M.} \&
  \bibinfo{author}{Sands, T.}
\newblock \bibinfo{title}{Lasers for {Satellite} {Uplinks} and {Downlinks}}.
\newblock \emph{\bibinfo{journal}{Sci}} \textbf{\bibinfo{volume}{3}},
  \bibinfo{pages}{4} (\bibinfo{year}{2021}).
\newblock \urlprefix\url{https://www.mdpi.com/2413-4155/3/1/4}.
\newblock \bibinfo{note}{Number: 1 Publisher: Multidisciplinary Digital
  Publishing Institute}.

\bibitem{seas_optical_2018}
\bibinfo{author}{Seas, A.~A.}, \bibinfo{author}{Robinson, B.},
  \bibinfo{author}{Shih, T.}, \bibinfo{author}{Khatri, F.} \&
  \bibinfo{author}{Brumfield, M.}
\newblock \bibinfo{title}{Optical {Communications} {Systems} for {NASA}'s
  {Human} {Space} {Flight} {Missions}} (\bibinfo{address}{Chania},
  \bibinfo{year}{2018}).
\newblock \urlprefix\url{https://ntrs.nasa.gov/citations/20180007086}.
\newblock \bibinfo{note}{NTRS Author Affiliations: NASA Goddard Space Flight
  Center, Massachusetts Inst. of Technology NTRS Report/Patent Number:
  GSFC-E-DAA-TN60456 NTRS Document ID: 20180007086 NTRS Research Center:
  Goddard Space Flight Center (GSFC)}.

\bibitem{xu_differential_2004}
\bibinfo{author}{Xu, C.}, \bibinfo{author}{Liu, X.} \& \bibinfo{author}{Wei,
  X.}
\newblock \bibinfo{title}{Differential phase-shift keying for high spectral
  efficiency optical transmissions}.
\newblock \emph{\bibinfo{journal}{IEEE Journal of Selected Topics in Quantum
  Electronics}} \textbf{\bibinfo{volume}{10}}, \bibinfo{pages}{281--293}
  (\bibinfo{year}{2004}).
\newblock \bibinfo{note}{Conference Name: IEEE Journal of Selected Topics in
  Quantum Electronics}.

\bibitem{toyoshima_recent_2021}
\bibinfo{author}{Toyoshima, M.}
\newblock \bibinfo{title}{Recent {Trends} in {Space} {Laser} {Communications}
  for {Small} {Satellites} and {Constellations}}.
\newblock \emph{\bibinfo{journal}{Journal of Lightwave Technology}}
  \textbf{\bibinfo{volume}{39}}, \bibinfo{pages}{693--699}
  (\bibinfo{year}{2021}).
\newblock \bibinfo{note}{Conference Name: Journal of Lightwave Technology}.

\bibitem{toyoshima_current_2015}
\bibinfo{author}{Toyoshima, M.} \emph{et~al.}
\newblock \bibinfo{title}{Current status of research and development on space
  laser communications technologies and future plans in {NICT}}.
\newblock In \emph{\bibinfo{booktitle}{2015 {IEEE} {International} {Conference}
  on {Space} {Optical} {Systems} and {Applications} ({ICSOS})}},
  \bibinfo{pages}{1--5} (\bibinfo{year}{2015}).

\bibitem{gottlieb_space_2019}
\bibinfo{author}{Gottlieb, J.}
\newblock \bibinfo{title}{Space {Colonization} and {Existential} {Risk}}.
\newblock \emph{\bibinfo{journal}{Journal of the American Philosophical
  Association}} \textbf{\bibinfo{volume}{5}}, \bibinfo{pages}{306--320}
  (\bibinfo{year}{2019}).
\newblock
  \urlprefix\url{https://www.cambridge.org/core/journals/journal-of-the-american-philosophical-association/article/abs/space-colonization-and-existential-risk/B82206D1268B2C9221EEA64B6CB14416}.
\newblock \bibinfo{note}{Publisher: Cambridge University Press}.

\bibitem{beckstead_how_2015}
\bibinfo{author}{Beckstead, N.}
\newblock \bibinfo{title}{How much could refuges help us recover from a global
  catastrophe?}
\newblock \emph{\bibinfo{journal}{Futures}} \textbf{\bibinfo{volume}{72}},
  \bibinfo{pages}{36--44} (\bibinfo{year}{2015}).
\newblock
  \urlprefix\url{https://www.sciencedirect.com/science/article/pii/S0016328714001888}.

\end{thebibliography}


\end{document}